\documentclass[twocolumn,english,amssymb,pra,superscriptaddress]{revtex4-1}
\usepackage{amsmath}
\usepackage{amssymb,amsmath}
\usepackage{graphicx}

\newcommand{\bra}[1]{\langle #1 |}
\newcommand{\ket}[1]{| #1 \rangle}
\newcommand{\tr}[1]{\textrm{Tr}\left[ #1 \right]}

\newcommand{\ts}{\textrm{s}}
\newcommand{\tf}{\textrm{f}}
\newcommand{\tin}{\alpha}
\newcommand{\opt}{\textrm{NLA}}
\newcommand{\var}[1]{\textrm{Var}\left(#1 \right)}
\newcommand{\mse}[1]{\textrm{MSE}\left(#1 \right)}

\begin{document}

\title{Phase Estimation of Coherent States with a Noiseless Linear Amplifier}

\author{Syed M Assad}
\email[E-mail: ]{cqtsma@gmail.com}
\author{Mark Bradshaw}
\author{Ping Koy Lam}

\affiliation{Centre for Quantum Computation and Communication
  Technology, Research School of Physics and Engineering, The
  Australian National University, Canberra ACT 2601, Australia}

\begin{abstract}
  Amplification of quantum states is inevitably accompanied with the
  introduction of noise at the output. For protocols that are
  probabilistic with heralded success, noiseless linear amplification
  in theory may still possible. When the protocol is successful, it
  can lead to an output that is a noiselessly amplified copy of the
  input.  When the protocol is unsuccessful, the output state is
  degraded and is usually discarded. Probabilistic protocols may improve the
  performance of some quantum information protocols, but not for
  metrology if the whole statistics is taken into
  consideration. We calculate the precision limits on estimating the
  phase of coherent states using a noiseless linear amplifier by
  computing its quantum Fisher information and we show that on average,
  the noiseless linear amplifier does not improve the phase
  estimate. We also discuss the case where abstention from measurement
  can reduce the cost for estimation.
\end{abstract}

\maketitle

\section{Introduction}
Quantum metrology is concerned with the measuring of a weak signal
with the best achievable precision by using a quantum probe. One
important example is in the detection of gravitational waves by
measuring the phase difference of light. It would be beneficial if we
could somehow amplify the signal prior to measurement. If the signal
is encoded as the amplitude $\alpha$ of a coherent state
$\ket{\alpha}$, a noiseless linear amplifier (NLA) can do just
that. An NLA with an amplification gain $g>1$ transforms the coherent
state $\ket{\alpha}$ to $\ket{g \alpha}$~\cite{Ralph2009}, thereby
amplifying the signal but not the noise. If this transformation can be
performed deterministically, we would obtain a more precise estimate
of the signal. Unfortunately, it is not possible to noiselessly
amplify a quantum state~\cite{Caves1982}. But an approximate version
of the NLA which works probabilistically is possible and has been
realised by several experimental
groups~\cite{Usuga2010,Xiang2010,Ferreyrol2010,Ferreyrol2011,Chrzanowski2014}.

\begin{figure}[!t]
\includegraphics[width=0.47\textwidth]{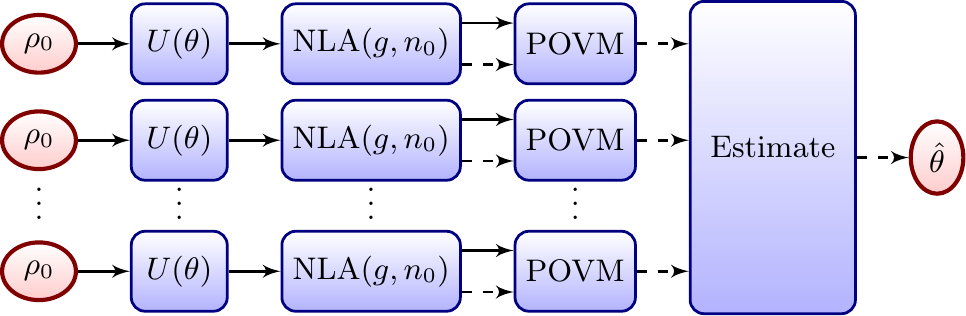}
\caption{{\bf Schematic of parameter estimation with an NLA.} Identical
  probes in some initial state $\rho_0$ undergoes an interaction
  $U(\theta)$. The probes are then individually amplified in an NLA
  with gain $g$, and maximum photon number $n_0$. The NLA outputs a
  projected quantum state (solid arrow) and a classical variable
  (dashed arrow) indicating successful or failed amplification. Based
  on the classical variable, an optimal POVM is chosen to measure the
  quantum state. The output of these measurements are used to obtain
  an estimate for $\theta$.}
    \label{fig:f0}
  \end{figure}

We investigate the precision of phase estimation of coherent states
using a probabilistic NLA as shown schematically in
Fig.~\ref{fig:f0}. When the NLA successfully amplifies a coherent
state, we are able to estimate the phase more precisely. However, when
the amplification fails, we obtain a worse estimate of the phase than
if we had not used the NLA. We show that on average, post-selecting
the successfully amplified events or using both successful and
unsuccessful events of the NLA does not improve the precision of phase
estimation. This is consistent with known results that by
post-selecting based on the measurement outcomes, probabilistic
metrology can result in improved quantum state estimation of the
post-selected sub-ensemble~\cite{Fiurasek2006,Gendra2012,Gendra2013},
but on average post-selection cannot increase
information~\cite{Tanaka2013,Ferrie2014,Combes2014,Zhang2015b}. However
with a different figure of merit, post-selection can help. This is the
case for state discrimination when a cost is assigned to wrong guesses
and for abstaining~\cite{Combes2015}. For our case, by assigning a
cost to rejecting a state and a cost for performing an estimator
measurement, then by post-selecting the successful outcome of the NLA
and only performing the estimator measurement on these, we can achieve
a desired precision at a lower cost.

\section{Phase estimation}
To quantify the precision of an estimate, we shall use the quantum
Fisher information~\cite{Helstrom1976,Holevo1982,Braunstein1994,Petz2011}. Given a sample of $m$ identical and independent states
$\rho_\theta$ that depend on some unknown parameter $\theta$ that we wish to
estimate, the quantum Cram\'{e}r Rao (QCR) bound states that the variance of an unbiased
estimator $\hat\theta$ is bounded by
\begin{align}
  \var{\hat{\theta}} \geq \frac{1}{ m J\left(\rho_\theta \right)}\;,
\end{align}
where $J\left(\rho_\theta \right)$ is the quantum Fisher information
\begin{align}
 J\left(\rho_\theta \right) =\tr{\rho_\theta \mathcal{L}^2}\;.
\end{align}
The symmetric logarithmic derivative $\mathcal L$ is some Hermitian
operator defined implicitly through
\begin{align}
 \dot\rho_\theta = \frac{1}{2} \left(\rho_\theta \mathcal L +
  \mathcal L
  \rho_\theta\right)\;,
\end{align}
where an overdot is used to indicate a derivative with respect to $\theta$. The QCR bound is asymptotically attainable when
$m\gg 1$~\cite{Barndorff-Nielsen2000}. A large Fisher information
allows for a more precise estimate of the parameter
$\theta$. Equivalently, a larger Fisher information allows a parameter
$\theta$ to be estimated to the same precision from a smaller
sample. For a pure state,
$\rho_\theta = \ket{\psi_\theta}\bra{\psi_\theta}$, we have
$\dot{\rho}_\theta = \rho_\theta \dot\rho_\theta + \dot\rho_\theta
\rho_\theta$ which indicates that we can take
$\mathcal L=2 \dot{\rho}_\theta$. This gives
$J\left(\rho_\theta\right)=4\tr{\rho_\theta \dot\rho_\theta^2} = 4
\left(\langle\dot \psi_\theta|\dot\psi_\theta \rangle + \langle
  \psi_\theta|\dot\psi_\theta \rangle ^2\right)$~\cite{Fujiwara1995}.

We apply the above formalism to an NLA. The NLA we consider is
implemented through a two outcome measurement device characterised by
a gain $g>1$ and maximum amplified photon
$n_0\in \mathbb{Z}^+$~\cite{Pandey2013,McMahon2014}. $n_0$ determines
how closely the successfully amplified output from this device resembles
the output from ideal NLA. A larger $n_0$ gives a more faithful
approximation at the expense of a lower probability of success. The
first measurement outcome corresponds to the operator
\begin{align}
  \label{eq:1}
  E_\ts =\sum_{n=0}^{n_0} g^{n-n_0} \ket{n}\bra{n} +\sum_{n=n_0+1}^{\infty} \ket{n}\bra{n}
\end{align}
which heralds a successful amplification event and projects the input
state $\rho_{\theta}$ to the state
$\rho_{\ts,\theta} = E_\ts \rho_{\theta} E_\ts/{\tr{\rho_{\theta}
    E_\ts^2}}$. The successful amplification event occurs with
probability $p_{\ts}=\tr{\rho_{\theta} E_\ts^2}$. The second
measurement outcome $E_\tf=\sqrt{1-E_\ts^2}$ corresponds to a failed
amplification event which projects the input state to
$\rho_{\tf,\theta} = E_\tf \rho_{\theta} E_\tf/{\tr{\rho_{\theta}
    E_\tf^2}}$ and occurs with probability
$p_{\tf}=\tr{ \rho_{\theta} E_\tf^2}$. We assume that $p_\ts$ and
$p_\tf$ do not depend on $\theta$ which is true for the state that we
shall consider later. From the
states $\rho_{\ts,\theta}$ we can construct $\hat\theta_\ts$, an
estimator of $\theta$, while from the states $\rho_{\tf,\theta}$, we
construct a second estimator $\hat\theta_\tf$. Combining these two
independent estimators, we arrive at a third estimator given by
$\hat{\theta}_\opt= \beta \hat\theta_\ts+(1-\beta) \hat\theta_\tf$.
The weight
\begin{align}
  \label{eq:beta}
  \beta=\frac{V_\tf}{V_\ts+V_\tf}\;,
\end{align}
where $V_\ts$ and $V_\tf$ denote the variances of $\hat\theta_\ts$ and
$\hat\theta_\tf$, is chosen to minimise the variance of
$\hat{\theta}_\opt$. The variances $V_\ts$ and $V_\tf$ depend on the
number of successful and failed amplification events denoted by
$n_\ts$ and $n_\tf$ respectively. Hence the weight $\beta$ is also a
function of number successfully amplified event $n_\ts$. The variance
of the estimator $\hat{\theta}_\opt$ given $n_\ts$ is
\begin{align}
  \var{\hat{\theta}_\opt |n_\ts}=\frac{1}{\frac{1}{V_\ts}+\frac{1}{V_\tf}}\geq
  \frac{1}{ \left(n_\ts J_\ts + n_\tf J_\tf \right)} \;,
\end{align}
using the notation $J_\ts=J\left( \rho_{\ts,\theta} \right)$ and $J_\tf=J\left( \rho_{\tf,\theta} \right)$.  We
define $J_\opt = \left(n_\ts J_\ts + n_\tf J_\tf\right)/m$, where
$m=n_\ts+n_\tf$ is the sample size. For a fixed $m$, $n_\ts$ follows a
binomial distribution with
$\Pr\left(n_\ts\right)=\binom{m}{n_\ts} p_\ts^{n_\ts}
p_\tf^{n_\tf}$. The unconditional variance of $\hat{\theta}_\opt$ is
then
\begin{equation}
\begin{split}
  \var{\hat{\theta}_\opt }&= \mathbb{E}_{n_\ts}\left[
  \var{\hat{\theta}_\opt |n_\ts} \right]  + \var{ \mathbb{E}\left[\hat{\theta}_\opt|n_\ts
  \right]}\\
&= \mathbb{E}_{n_\ts}\left[
  \var{\hat{\theta}_\opt |n_\ts} \right]\;,
\end{split}
\end{equation}
since $n_\ts$ does not depend on $\theta$. For large $m$, $n_\ts/m
\rightarrow p_\ts$ and $n_\tf/m
\rightarrow p_\tf$ so that 
$J_\opt  \rightarrow p_\ts J_\ts+p_\tf J_\tf$~\cite{Zhang2015b}.

\begin{figure}[!t]
\includegraphics{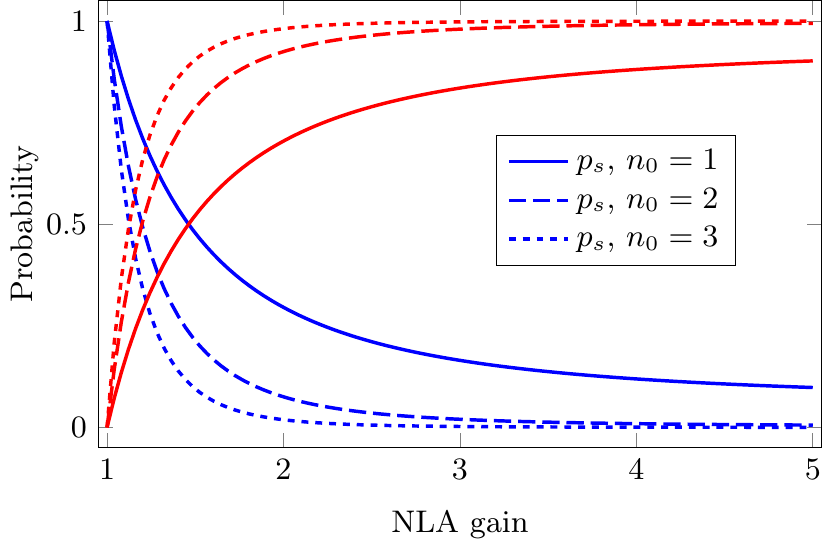}
\caption{{\bf Probability of successful (blue, decreasing) and failed
    (red, increasing) NLA amplification versus NLA gain.} The blue and
  red lines add up to 1. Input state has amplitude $r=0.25$.}
    \label{fig:ps}
\end{figure}

We consider a coherent input state
$\rho_\tin=\ket{\alpha}\bra{\alpha}$ with $\alpha=r e^{i \theta}$,
where the amplitude $r$ is known and whose phase $\theta$ we wish to
estimate. The quantum Fisher information for $\rho_\tin$ is
$J_\tin=4r^2$~\cite{Bagan2008,Aspachs2009,Pinel2013}. Applying the NLA
on the state $\ket \alpha$, we get one of the two outputs
\begin{equation}
\begin{split}
  \ket{\psi_\ts}&=E_\ts \ket{\alpha}\\
                &=                 \frac{\exp\left(\frac{-r^2}{2}\right)}{\sqrt{p_\ts(r)}}\left(\sum_{n=0}^{n_0}  \ket{n}
                 \frac{(g\alpha)^n}{\sqrt{n!}g^{n_0}}
                 +\sum_{n=n_0+1}^{\infty}
                 \ket{n}\frac{\alpha^n}{\sqrt{n!}}\right)
\end{split}
\end{equation}
or
\begin{equation}
\begin{split}
  \ket{\psi_\tf}&=E_\tf \ket{\alpha}\\
                &=                 \frac{\exp\left(\frac{-r^2}{2}\right)}{\sqrt{p_\tf(r)}}\sum_{n=0}^{n_0}  \ket{n}
                 \sqrt {1-\frac{g^{2n}}{g^{2n_0}} }\frac{\alpha^n}{\sqrt{n!}}\
\end{split}
\end{equation}
with probabilities
\begin{align}
  p_\ts&=\exp \left(-r^2\right) \left(\sum_{n=0}^{n_0} \frac{g^n
         r^{2n}}{n! g^{n_0}}  +\sum_{n=n_0+1}^{\infty} \frac{r^{2 n}}{n!}\right)\\
  p_\tf&=\exp\left(-r^2\right) \sum_{n=0}^{n_0} \left( 1-\frac{g^{2n}}{g^{2n_0}}\right) \frac{r^{2 n}}{n!}\;
\end{align}
that do not depend on $\theta$. The probability of success and failure are
plotted in Fig.~\ref{fig:ps} for $r=0.25$. As $n_0$
increases, we get a better approximation to the ideal NLA
transformation but at the expense of a lower probability of success.
Differentiating the outputs, we get the unnormalised states
\begin{align}
  \ket{\dot\psi_\ts}&=                 \frac{\exp\left(\frac{-r^2}{2}\right)}{\sqrt{p_\ts(r)}}\left(\sum_{n=0}^{n_0}  \ket{n}
                 \frac{i\,(g\alpha)^n n}{\sqrt{n!}g^{n_0}}
                 +\sum_{n=n_0+1}^{\infty}
                 \ket{n}\frac{i\,\alpha^n n}{\sqrt{n!}}\right)\\
  \ket{\dot\psi_\tf}&=  \frac{\exp\left(\frac{-r^2}{2}\right)}{\sqrt{p_\tf(r)}}\sum_{n=0}^{n_0}  \ket{n}
                 \sqrt{ 1-\frac{g^{2n}}{g^{2n_0}} }\frac{i\, \alpha^n n}{\sqrt{n!}}
\end{align}
with which we can compute $J_\ts$ and $J_\tf$.

\begin{figure}[!t]
\includegraphics{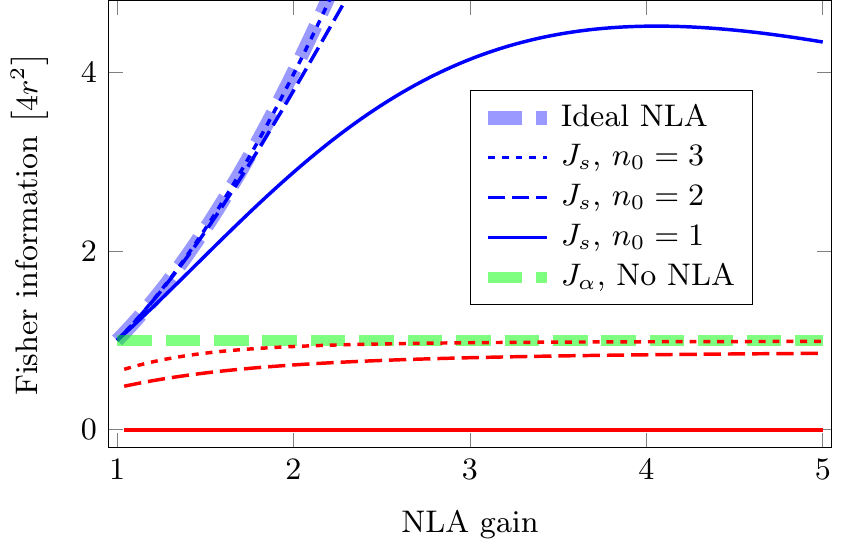}
\caption{{\bf Fisher information of successful and failed
    amplification events versus NLA gain.} $J_\ts$ (blue lines) is
  the Fisher information when the NLA successfully amplify the
  state. In these case, the Fisher information is higher than the
  Fisher information without the NLA, $J_\alpha$ (green line).
  $J_\tf$ (red lines) is the Fisher information when the NLA failed
  to amplify the state. For these case, $J_\tf$ is lesser than
  $J_\alpha$.  $J_\textrm{ideal}$ (thick blue line) is the Fisher
  information of the state $\ket{g \alpha}$ that one will obtain from
  a successful NLA with a large $n_0$. Input state has amplitude
  $r=0.25$ and the Fisher information are normalised such that
  $J_\tin=1$. }
    \label{fig:1}
\end{figure}

\begin{figure}[!t]
\includegraphics{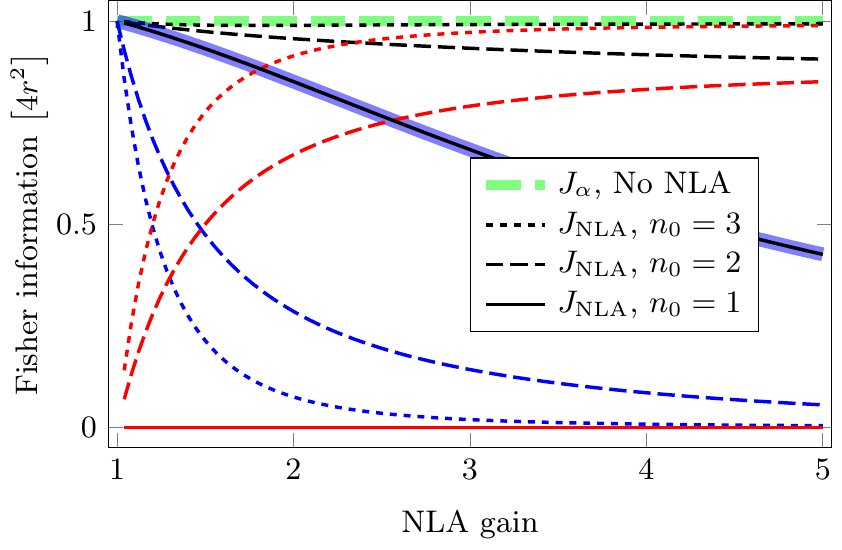}
\caption{{\bf Breakdown of the Fisher informations of the NLA when
    $p_\ts=n_\ts/m$ versus NLA gain.} The scaled information from the
  successful NLA events $p_\ts J_\ts$ (blue lines) decreases as the
  NLA gains increases due to the low probability of success, while
  scaled information from the failed NLA events $p_\tf J_\tf$ (red
  lines) increases with higher gain. Their sum gives the net
  information $J_\opt$ (black lines) which is always lower than the
  Fisher information one gets without the NLA $J_\alpha$ (green line).
  Input state has amplitude $r=0.25$ and the Fisher information are
  normalised such that $J_\tin=1$. }
    \label{fig:1b}
  \end{figure}

We plot the Fisher information $J_\tin$, $J_\ts$ and $J_\tf$ as a
function of NLA gain in Fig.~\ref{fig:1}. The successfully amplified
states $\ket{\psi_\ts}$ have higher Fisher information compared to the
input coherent states, while the failure states $\ket{\psi_\tf}$ have
a lower Fisher information. Hence, we can probabilistically get a
higher information when the amplification succeed. For $n_0=1$, the
states $\ket{\psi_\tf}$ carries no information about the phase
$\theta$.  In Fig.~\ref{fig:1b}, we plot the Fisher information scaled
by their respective probabilities. We see that $p_\ts J_\ts$ and
$p_\tf J_\tf$ are both lower than $J_\alpha$. Their sum $J_\opt$, is
also always lower than the Fisher information without using an
NLA. This demonstrates the fact that doing a post-selection cannot
increase information~\cite{Tanaka2013,Combes2014,Zhang2015b}. From
Fig.~\ref{fig:ps}, we see that when $g$ increases, there is a much
higher probability for the amplification to fail.  For $n_0>1$, this
results in more net information gained from the failed amplification
events than the successfully amplified events at high $g$.

\begin{figure}[!t]
\includegraphics{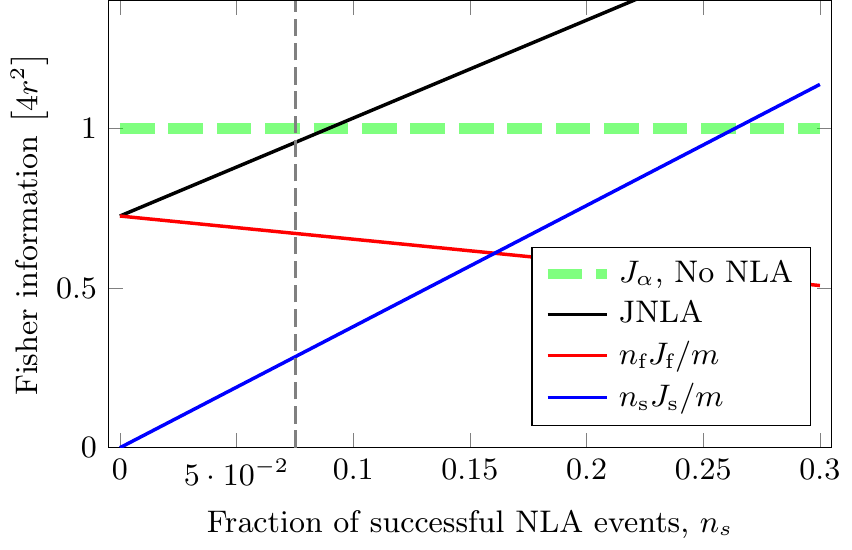}
\caption{{\bf Breakdown of the Fisher informations of the NLA versus
    fraction of successfully amplified states $n_\ts/m$.}  Fisher
  information from the successfully amplified states (blue line)
  increases when the fraction of successful NLA events increases while
  the Fisher information from the unsuccessfully amplified states (red
  line) decreases. The net Fisher information (black line) can be
  higher than the Fisher information one gets without using the NLA
  $J_\alpha$ (green line) when $n_\ts$ is large enough. The probability of this
  happening is small when the sample size $m$ is large. The vertical
  line denotes the most likely $n_\ts=m p_\ts$. On this line, $J_\opt$
  is less than $J_\alpha$.  Input state has amplitude $r=0.25$. The
  NLA has a gain $g=2$ and $n_0=2$. The Fisher information are
  normalised such that $J_\tin=1$. }
    \label{fig:2}
\end{figure}

In Fig.~\ref{fig:2}, we fix the NLA gain $g=2$, and plot the
Fisher information $J_\opt$ as a function of the fraction
of successfully amplified states $n_\ts/m$. We see that as
$n_\ts$ increases, $J_\opt$ increases and eventually becomes larger
than $J_\alpha$. However, the probability to get a large enough $n_\ts$ is small
when the sample size $m$ is large. For example, for $m=1000$, we need
$n_\ts>89$ before $J_\opt>J_\alpha$. The probability for this is only $4.68\%$. The vertical line indicates the mean value of
$n_\ts/m = p_\ts$. At this value, $J_\opt$ is less than $J_\alpha$.

\section{Simulations with finite sample}

For small $\theta$ and pure state $\rho_\theta$, the QCR bound can be attained by
measuring the observable $\mathcal C=\lambda^2 \mathcal{L}$ where $\mathcal L=2\left(\ket{\psi_0}\bra{\dot\psi_0}
  +\ket{\dot{\psi}_0}\bra{\psi_0}\right)$ has rank at most
two and $\lambda^2=1/\left(4\tr{\rho_0 \dot\rho_0^2}\right)$. The
estimator obtained through $\mathcal{C}$ has moments
\begin{align}
  \label{eq:5}
  \tr{ \rho_{\theta} \mathcal C}  &= \theta +O(\theta^2)\\
  \tr{ \rho_{\theta} \mathcal C^2} &= \lambda^2  +O(\theta^2)
\end{align}
which verify that $\mathcal C$ is an unbiased estimator of $\theta$ achieving the QCR
bound. The observable $\mathcal C$ has zero trace and spectral
decomposition $\ket{c_+}\lambda \bra{c_+} -\ket{c_-}\lambda
\bra{c_-}$ where $\ket{c_+}$ and $\ket{c_-}$ are orthonormal
vectors. Given $m$ trials, the probability of obtaining $n_+$ positive
outcomes and $n_-$ negative
outcomes follows a multinomial distribution
$\Pr\left({n_+,n_-}\right)=\frac{m}{n_+! n_-! n_0!}p_{+}^{n_+}p_{-}^{n_-}p_0^{n_0}$ where
$n_0=m-n_--n_+$ and the event probabilities
\begin{equation}
\begin{split}
  \label{eq:3}
  p_\pm &= \left<c_\pm|\rho_\theta|c_\pm \right>\\
  &=\frac{1}{2}\left(1 \pm \frac{\theta}{\lambda} \right)+O\left(\theta^2\right)\;
\end{split}
\end{equation}
and $p_0=1-p_--p_+$.

\begin{figure}[!t]
\includegraphics{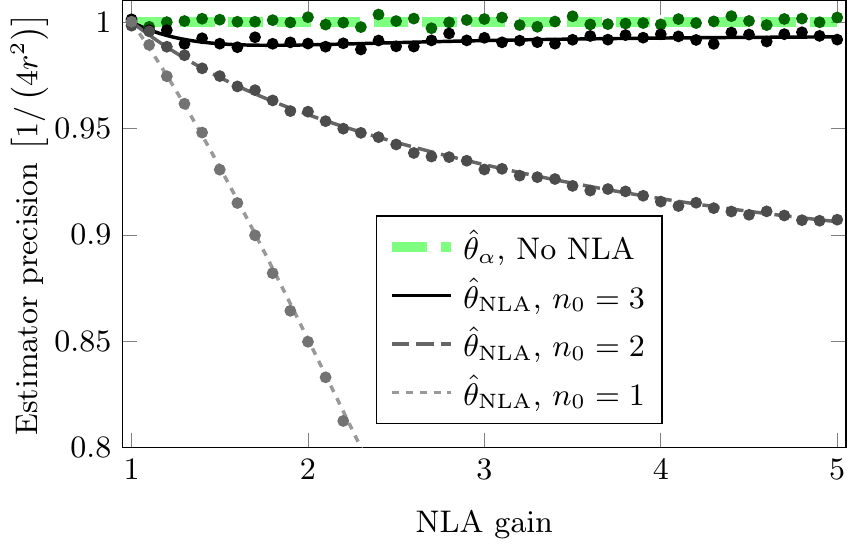}
\caption{{\bf Simulation of estimator precision for NLA versus NLA
    gain.} The measurement precision using the NLA (black lines) is
  always worse than without using the NLA (green line). Increasing the
  NLA gain does not make it better. Estimates were obtained from a sample
  size of $m=1000$ input states with $r=0.25$ and
  $\theta_\textrm{true}=0.01$. The datapoints were obtained by
  measuring the precision from $10^6$ estimation runs. The lines are
  the asymptotic theoretical precisions conditioned on the most likely
  values of $n_s$ and $n_f$. The precision is scaled such that
  $\textrm{Precision}(\hat{\theta}_\tin)=1$.}
    \label{fig:sim}
\end{figure}

For coherent states without the NLA, $\lambda_\alpha=1/(2 r)$ and
$\mathcal C_\alpha=\mathcal L_\alpha/(4 r^2)$ with 
\begin{align}
  \label{eq:6}
  \tr{\mathcal C_\tin \rho_{\tin}}  &= \theta +O(\theta^2)\\
  \tr{\mathcal C_\tin^2 \rho_{\tin}} &= \frac{1}{4r^2}  +O(\theta^2)
\end{align}
is an optimal unbiased estimator of $\theta$. For $m$
measurements, the counts
$n_{\alpha +}$ and $n_{\alpha -}$ follows a multinomial
distribution with $m$ trials and event probabilities $p_{\alpha\pm}=
\left<c_{\alpha\pm}|\rho_\theta|c_{\alpha\pm} \right> $. $\mathcal
C_\alpha$ is the maximum likelihood estimator giving an
estimate~\cite{Rice2006}
\begin{align}
  \label{eq:7}
  \hat{\theta}_\alpha=\frac{n_{\alpha +}-n_{\alpha -}}{n_{\alpha +}+n_{\alpha -}}\lambda_\alpha\;.
\end{align}
The estimate obtained from the NLA can be viewed as an estimate
obtained from a five outcome POVM
$\left\{E_{\ts \pm}^2,E_{\tf \pm}^2,E_0^2 \right\}$, where
$E_{\ts\pm}^2= E_\ts \ket{c_{\ts\pm}} \bra{c_{\ts\pm}} E_\ts$,
$E_{\tf\pm}^2= E_\tf \ket{c_{\tf\pm}} \bra{c_{\tf\pm}} E_\tf$ and
$E_0^2=1-E_{\ts+}^2-E_{\ts-}^2-E_{\tf+}^2-E_{\tf-}^2$.
The vectors $\ket{c_{\ts\pm}}$ and $\ket{c_{\tf\pm}}$ are the eigenvectors of the
observable $\mathcal C_\ts$ and $\mathcal C_\tf$ with corresponding eigenvalues $\lambda_\ts$
and $\lambda_\tf$ for optimal estimation with the input states $\ket{\psi_\ts}$ and $\ket{\psi_\tf}$. Given $m$ measurements, the count
rates $n_{\ts \pm}$ and $n_{\tf \pm}$ follows a multinomial
distribution with $m$ trials and event probabilities
$p_{\ts\pm}=\tr{\rho_\tin E_{\ts\pm}^2}$ and
$p_{\tf\pm}=\tr{\rho_\tin E_{\tf\pm}^2}$. Given these counts, the
maximum likelihood estimate for $\theta$ is constructed by~\cite{Rice2006}
\begin{align}
  \label{eq:4}
  \hat{\theta}_\opt = \frac{n_\ts \lambda_\tf^2}{n_\ts
  \lambda_\tf^2+n_\tf \lambda_\ts^2} \hat{\theta}_\ts + \frac{n_\tf \lambda_\ts^2}{n_\ts
  \lambda_\tf^2+n_\tf \lambda_\ts^2} \hat{\theta}_\tf\;, 
\end{align}
which is consistent with Eq.~(\ref{eq:beta}) and where
$n_\ts=n_{\ts+}+n_{\ts-}$ and $n_\tf=n_{\tf+}+n_{\tf-}$. The
intermediate estimators are
$\hat\theta_\ts=(n_{\ts+}-n_{\ts-})\lambda_\ts/n_\ts$ and
$\hat\theta_\tf=(n_{\tf+}-n_{\tf-})\lambda_\tf/n_\tf$. We plot the
precision of the estimators $\hat{\theta}_\tin$ and
$\hat{\theta}_\opt$ defined by
\begin{align}
  \label{eq:8}
  \textrm{Precision}\left(\hat{\theta} \right) =\frac{1}{m\,\mse{\hat{\theta}}}
\end{align}
in Fig.~\ref{fig:sim}, where the mean square error (MSE)
of an estimator $\hat{\theta}$ is
\begin{align}
  \label{eq:9}
\mse{\hat\theta}=\mathbb{E}_{\hat{\theta}}\left[(\hat\theta - \theta_\textrm{true})^2 \right]\geq \var{\hat\theta}\;.
\end{align}
Here $\theta_\textrm{true}$ is the true value of the parameter
$\theta$. From Fig.~\ref{fig:sim}, we see that on average, the NLA does not increase the
precision for phase estimation.

\section{Discussion}
The NLA is well suited for some tasks where all that matters are the
successfully amplified states and when the probability of success does
not matter, such as in probabilistic entanglement distillation and
quantum key distribution~\cite{Chrzanowski2014}. In a phase estimation
problem, if the figure of merit is the precision from a given number
of sample, then as to be expected, using the NLA does not offer any advantage for phase
estimation when compared to the optimal phase estimation
scheme.

However with different figure of merits, using an NLA and
post-selecting only successfully amplified events can help in
metrology. Suppose we associate a cost $x$ for acquiring a sample, $y$
for direct measurement of an estimator observable from each sample and
$z$ for applying a noiseless linear amplification on a sample, and our
objective is to minimize the cost for obtaining an estimate for
$\theta$ to a specified precision $\epsilon$. In order to achieve the
specified precision without using the NLA, we would need to perform an
estimate on $m_\alpha=\epsilon/J_\alpha$ samples. The total cost is
then $\epsilon(x+y)/J_\alpha$. With the NLA, and performing an
estimate only when the NLA heralds a successful amplification event,
we now need to perform an estimate on only $m_\ts=\epsilon/J_\ts$
samples. Since $J_\ts > J_\alpha$, each measurement gives more
information and so we need less estimator measurements compared to
estimating without the NLA. However the total number of samples we
need to acquire increases because some samples were discarded when the
NLA did not herald a successful amplification. We now need on average
a total of $m_\ts /p_\ts$ samples and the total cost of the estimate
would be
$\epsilon\left(x+z+p_\ts y \right)/\left(p_\ts J_\ts \right)$. In
conventional metrology, the cost $y$ assigned to measuring an estimator
observable is zero, and since $p_\ts J_\ts < J_\alpha$, the cost from
the post-selection strategy will always be higher than without using
the NLA. In this case, post-selection does not help. However if $y$ is
non-zero, then the total cost of using the NLA and performing
post-selection can be lesser than a direct measurement on all samples. This is true
when
\begin{align}
y > \frac{ \left(J_\alpha - p_\ts J_\ts\right)x +J_\alpha z}{ p_\ts\left(J_\ts-J_\alpha \right)}\;.
\end{align}
In this case, the better strategy would be to abstain from measuring the
sample whenever the NLA fails to amplify.
\\
\\
\textbf{Acknowledgements} This research is supported by the
Australian Research Council (ARC) under the Centre of Excellence for
Quantum Computation and Communication Technology (CE110001027)


\begin{thebibliography}{100}

\bibitem{Ralph2009}
T.~C. Ralph and A.~P. Lund.
\newblock Nondeterministic noiseless linear amplification of quantum systems.
\newblock {\em AIP Conference Proceedings}, 1110(1):155--160, 2009.

\bibitem{Caves1982}
Carlton Caves.
\newblock Quantum limits on noise in linear amplifiers.
\newblock {\em Phys. Rev. D}, 26(8):1817--1839, Oct 1982.

\bibitem{Usuga2010}
Mario~A. Usuga, Christian~R. Müller, Christoffer Wittmann, Petr Marek, Radim
  Filip, Christoph Marquardt, Gerd Leuchs, and Ulrik~L. Andersen.
\newblock Noise-powered probabilistic concentration of phase information.
\newblock {\em Nat Phys}, 6(10):767--771, Oct 2010.

\bibitem{Xiang2010}
G.~Y. Xiang, T.~C. Ralph, A.~P. Lund, N.~Walk, and G.~J. Pryde.
\newblock Heralded noiseless linear amplification and distillation of
  entanglement.
\newblock {\em Nature Photonics}, 4(5):316--319, Mar 2010.

\bibitem{Ferreyrol2010}
Franck Ferreyrol, Marco Barbieri, R{\'e}mi Blandino, Simon Fossier, Rosa
  Tualle-Brouri, and Philippe Grangier.
\newblock Implementation of a nondeterministic optical noiseless amplifier.
\newblock {\em Phys. Rev. Lett.}, 104(12):123603, 2010.

\bibitem{Ferreyrol2011}
Franck Ferreyrol, R{\'e}mi Blandino, Marco Barbieri, Rosa Tualle-Brouri, and
  Philippe Grangier.
\newblock Experimental realization of a nondeterministic optical noiseless
  amplifier.
\newblock {\em Phys. Rev. A}, 83(6):063801, 2011.

\bibitem{Chrzanowski2014}
Helen~M. Chrzanowski, Nathan Walk, Syed~M. Assad, Jiri Janousek, Sara Hosseini,
  Timothy~C. Ralph, Thomas Symul, and Ping~Koy Lam.
\newblock Measurement-based noiseless linear amplification for quantum
  communication.
\newblock {\em Nature Photonics}, 8(4):333--338, Mar 2014.

\bibitem{Fiurasek2006}
J.~Fiurasek.
\newblock Optimal probabilistic estimation of quantum states.
\newblock {\em New J. Phys.}, 8(9):192–192, Sep 2006.

\bibitem{Gendra2012}
Bernat Gendra, Elio Ronco-Bonvehi, John Calsamiglia, Ramon Mu{\~n}oz-Tapia, and
  Emilio Bagan.
\newblock Beating noise with abstention in state estimation.
\newblock {\em New Journal of Physics}, 14(10):105015, Oct 2012.

\bibitem{Gendra2013}
B.~Gendra, E.~Ronco-Bonvehi, J.~Calsamiglia, R.~Muñoz-Tapia, and E.~Bagan.
\newblock Quantum metrology assisted by abstention.
\newblock {\em Physical Review Letters}, 110(10), Mar 2013.

\bibitem{Tanaka2013}
Saki Tanaka and Naoki Yamamoto.
\newblock Information amplification via postselection: A parameter-estimation
  perspective.
\newblock {\em Phys. Rev. A}, 88:042116, Oct 2013.

\bibitem{Ferrie2014}
Christopher Ferrie and Joshua Combes.
\newblock Weak value amplification is suboptimal for estimation and detection.
\newblock {\em Phys. Rev. Lett.}, 112(4), Jan 2014.

\bibitem{Combes2014}
Joshua Combes, Christopher Ferrie, Zhang Jiang, and Carlton~M. Caves.
\newblock Quantum limits on postselected, probabilistic quantum metrology.
\newblock {\em Phys. Rev. A}, 89:052117, May 2014.

\bibitem{Zhang2015b}
Lijian Zhang, Animesh Datta, and Ian~A. Walmsley.
\newblock Precision metrology using weak measurements.
\newblock {\em Phys. Rev. Lett.}, 114(21), May 2015.

\bibitem{Combes2015}
Joshua Combes and Christopher Ferrie.
\newblock Cost of postselection in decision theory.
\newblock {\em Phys. Rev. A}, 92(2), Aug 2015.

\bibitem{Helstrom1976}
Carl~W Helstrom.
\newblock {\em Quantum detection and estimation theory}.
\newblock Academic press, 1976.

\bibitem{Holevo1982}
A~Holevo.
\newblock {\em Probabilistic and Statistical Aspects of Quantum Theory}.
\newblock North-Holland, Amsterdam, 1982.

\bibitem{Braunstein1994}
Samuel~L. Braunstein and Carlton~M. Caves.
\newblock Statistical distance and the geometry of quantum states.
\newblock {\em Phys. Rev. Lett.}, 72(22):3439--3443, May 1994.

\bibitem{Petz2011}
D.~Petz and C.~Ghinea.
\newblock {\em Introduction to Quantum Fisher Information}, chapter~15, pages
  261--281.
\newblock World Scientific, Jan 2011.

\bibitem{Barndorff-Nielsen2000}
O~E Barndorff-Nielsen and R~D Gill.
\newblock Fisher information in quantum statistics.
\newblock {\em Journal of Physics A: Mathematical and General},
  33(24):4481–4490, Jun 2000.

\bibitem{Fujiwara1995}
Akio Fujiwara and Hiroshi Nagaoka.
\newblock Quantum fisher metric and estimation for pure state models.
\newblock {\em Phys. Lett. A}, 201(2-3):119–124, May 1995.

\bibitem{Pandey2013}
Shashank Pandey, Zhang Jiang, Joshua Combes, and Carlton~M. Caves.
\newblock Quantum limits on probabilistic amplifiers.
\newblock {\em Phys. Rev. A}, 88(3), Sep 2013.

\bibitem{McMahon2014}
N.~A. McMahon, A.~P. Lund, and T.~C. Ralph.
\newblock Optimal architecture for a nondeterministic noiseless linear
  amplifier.
\newblock {\em Phys. Rev. A}, 89:023846, Feb 2014.

\bibitem{Bagan2008}
E.~Bagan, A.~Monras, and R.~Mu{\~n}oz-Tapia.
\newblock Phase variance of squeezed vacuum states.
\newblock {\em Phys. Rev. A}, 78:043829, Oct 2008.

\bibitem{Aspachs2009}
M.~Aspachs, J.~Calsamiglia, R.~Mu{\~n}oz-Tapia, and E.~Bagan.
\newblock Phase estimation for thermal {Gauss}ian states.
\newblock {\em Phys. Rev. A}, 79(3), Mar 2009.

\bibitem{Pinel2013}
O.~Pinel, P.~Jian, N.~Treps, C.~Fabre, and D.~Braun.
\newblock Quantum parameter estimation using general single-mode gaussian
  states.
\newblock {\em Phys. Rev. A}, 88:040102, Oct 2013.

\bibitem{Rice2006}
John Rice.
\newblock {\em Mathematical statistics and data analysis}.
\newblock Nelson Education, 2006.

\end{thebibliography}
\end{document}